\pdfoutput=1
\documentclass[11pt]{article}
\usepackage{subcaption}
\usepackage{graphicx}
\usepackage{xcolor}
\usepackage{amsmath}

\usepackage[]{ACL2023}

\usepackage{times}
\usepackage{latexsym}

\usepackage{float}

\usepackage[T1]{fontenc}
\usepackage[utf8]{inputenc}

\usepackage{microtype}

\usepackage{inconsolata}
\usepackage{epigraph}

\usepackage{listings}
\usepackage{xcolor}

\title{The Turing Test is More Relevant Than Ever}

\author{Avraham Rahimov, Orel Zamler, and Amos Azaria\\
School of Computer Science \\
  Ariel University, Israel
  }

\begin{document}
\maketitle

\epigraph{\textit{"A computer would deserve to be called intelligent if it could deceive a human into believing that it was human."}}
{Alan Turing, 1950}

\begin{abstract}
The Turing Test, first proposed by Alan Turing in 1950, has historically served as a benchmark for evaluating artificial intelligence (AI). However, Since the release of ELIZA in 1966, and particularly with recent advancements in large language models (LLMs), AI has been claimed to pass the Turing Test. Furthermore, criticism argues that the Turing Test primarily assesses deceptive mimicry rather than genuine intelligence, prompting the continuous emergence of alternative benchmarks. This study argues against discarding the Turing Test, proposing instead using more refined versions of it, for example, by interacting simultaneously with both an AI and human candidate to determine who is who, allowing a longer interaction duration, access to the Internet and other AIs, using experienced people as evaluators, etc.

Through systematic experimentation using a web-based platform, we demonstrate that richer, contextually structured testing environments significantly enhance participants' ability to differentiate between AI and human interactions. Namely, we show that, while an off-the-shelf LLM can pass some version of a Turing Test, it fails to do so when faced with a more robust version.  Our findings highlight that the Turing Test remains an important and effective method for evaluating AI, provided it continues to adapt as AI technology advances. Additionally, the structured data gathered from these improved interactions provides valuable insights into what humans expect from truly intelligent AI systems.
\end{abstract}

\section{Introduction}

The Turing Test, proposed by Alan Turing in 1950~\citep{turing1950computing}, has historically served as a foundational benchmark for assessing artificial intelligence (AI). Over the decades, several AI systems—most notably ELIZA~\citep{weizenbaum1966eliza} and Eugene Goostman~\citep{warwick2016can}—have claimed to pass variations of this test, often sparking debate over its adequacy as a measure of genuine intelligence. Recently, the consensus has shifted: modern large language models (LLMs) such as GPT-4~\citep{bubeck2023sparks} convincingly pass traditional forms of the Turing Test, diminishing its perceived relevance. Critics argue that this benchmark is primarily focused on deception rather than meaningful intelligence, prompting researchers to continually propose new benchmarks which, though initially challenging, are often surpassed within months~\citep{srivastava2022beyond}.

Despite such criticisms, this paper argues that dismissing the Turing Test as outdated overlooks its significant potential as an ultimate evaluation of general intelligence, provided it is adapted to contemporary AI advancements. Rather than abandoning it, we propose that the Turing Test can—and should—be updated. Modern adaptations could include extended interaction times, engagement with domain experts as evaluators, enabling real-world interactions (such as placing online orders, composing a presentation, building websites, and creating videos), or incorporating audio and video communication. Moreover, contemporary versions of the test might allow both AI and human participants to leverage the Internet and even collaborate with other AIs. The human responder could be an expert in their field, for example, an expert software programmer. Crucially, the test should strongly incentivize human testers to pose meaningful challenges and human responders to convincingly demonstrate their authenticity.

If an AI consistently remains indistinguishable from a human across diverse, extended, and complex interactions, this would provide robust evidence of achieving genuine human-level intelligence (albeit, in the virtual world). Furthermore, data collected from these enriched interactions would offer invaluable insights, accurately reflecting human expectations and standards for general intelligence.

To systematically explore these issues, we introduce a modern web-based platform designed to examine two different environment settings of the Turing test, which we refer to as simple and enhanced.

In the simple variant, which is based on the experiment conducted by \cite{biever2023chatgpt}, a human interacts briefly with a single candidate and must determine if she was conversing with an AI or a human. However, in the enhanced variant, we take several measures to ensure a more robust test, with the most significant difference being that the participants interact using a dual-chat interface. That is, the human participants (testers) simultaneously interact with both a human (responder) and an AI chatbot without knowing who is who. This allows the tester to compare the responses obtained from both parties when deciding who is human and who is an AI chatbot. 

The primary objectives of this research are threefold. First, we aim to establish a standardized and reproducible environment for conducting Turing Test experiments. Second, we investigate how environmental factors, such as participant selection, conversation duration, and interface design, influence the overall human-AI interaction, rather than just AI performance. Finally, we assess the impact of engagement strategies and incentives on user participation and decision-making, while also comparing the effectiveness of simple versus enhanced Turing Tests in differentiating between AI and human intelligence. To the best of our knowledge, this is the first work to compare the performance (with respect to passing the Turing test) of identical models, in two different environments.

By emphasizing the key differences between simple and enhanced Turing Tests, this study demonstrated the necessity of refining AI evaluation methodologies rather than abandoning the Turing Test altogether. The experimental environment, participant composition, and testing duration all play essential roles in determining evaluation outcomes. Our findings contribute to ongoing discussions in AI research by demonstrating that while AI may easily deceive users in simplistic settings, more comprehensive and structured tests reveal deeper limitations, reinforcing the relevance of the Turing Test as an important benchmark for evaluating general intelligence. Indeed, now that modern LLMs excel at generating convincing language, achieving high performance on a well-designed Turing Test requires AI to demonstrate true human-level intelligence across diverse tasks, rather than merely mimicking human-like conversation.

\section{Related Work}

The Turing Test \cite{turing1950computing} was designed as an operational criterion for machine intelligence. In its original formulation, commonly referred to as the ``imitation game'', a human judge (interrogator) engages in a text-based conversation with both a human and a machine simultaneously, without knowing which is which. If the judge cannot reliably distinguish between them, the machine is said to exhibit intelligent behavior. While Turing did not specify a strict numerical threshold for passing the test, later interpretations often cite a benchmark of around 66.7-70\%, i.e., if only two-thirds or less, of human testers correctly identify the machine as non-human, the machine is considered to have passed the test. In addition, \cite{jones2025large} propose that passing the Turing test requires that success rate will be non-statistically significant lower than 50\%.

The Loebner Prize competition was a practical instantiation of Turing’s original ideas \cite{bradevsko2012survey}, held annually from 1991 until its discontinuation in 2019. A 2012 survey of chatbot systems developed for the competition highlights that, at the time, many chatbots were able to superficially mimic human conversation but largely relied on pre-scripted responses and heuristic rules rather than genuine contextual understanding. As a result, these systems often failed to maintain sustained coherence or handle open-ended or ambiguous queries. The competition has drawn substantial criticism over the years, including concerns that it prioritized deception over intelligence, encouraged superficial ELIZA-style gimmicks, and relied on untrained judges making quick decisions under overly constrained conditions. While such critiques were valid, especially given the technical limitations of the time, the discontinuation of the prize coincided with the emergence of LLMs, which are capable of generating contextually rich, coherent, context-sensitive responses across a wide range of topics. As a result, the key challenge for AI is no longer simply producing human-like conversation, which current LLMs can achieve to some extent, but demonstrating deeper forms of intelligence, such as reasoning, abstraction, adaptability, and the ability to perform complex tasks, which the original Turing Test implicitly aimed to evaluate.

% The Turing Test, proposed by Alan Turing in 1950, remains one of the most debated measures of artificial intelligence (AI). As large language models (LLMs) such as GPT-4 advance, researchers have conducted various experiments to evaluate their ability to mimic human conversation. However, the effectiveness of these tests depends heavily on the experimental environment, as an incomplete or flawed setup may lead to misleading conclusions about an AI’s ability to pass the Turing Test. In this section, we analyze three studies that have examined AI-human indistinguishability and highlight key shortcomings in their methodologies.

% verify and insert: Turing suggested the threshold to be 66.7\%, that is, if only two thirds of the human testers correctly detect who is human and who is a machine, the machine is said to pass the test.

%\subsection{Weaknesses in "Human or Not? A Gamified Approach to the Turing Test"}

Indeed, several recent studies have attempted to evaluate LLMs through variations of the Turing Test. In~\cite{jannai2023human}, Jannai et al. conducted a large-scale study, by introducing the online game {\it ``Human or Not?''}. In this game, over 1.5 million participants engaged in short conversations with either a human or an AI model, after which they had to indicate whether they were interacting with another human or with an AI. When interacting with an AI, only 62\% of the humans correctly identified it as an AI. Thus, Jannai et al. conclude that their prompt engineered LLM has passed the Turing test. However, despite its large-scale participation, the experimental design introduces several critical flaws.

First, the interaction was conducted with only a single chat, so one could not compare the behavior of the AI to that of a real human. In addition, the participants were not assigned specific roles, with one being a tester and the other being a responder, and instead, the humans and the AI also played the role of a tester and also of a responder. Furthermore, the participants were not provided clear instructions, so some participants took the role of an AI, and attempted to answer as an AI would answer instead of acting like a human. Additionally, each interaction lasted only two minutes, with an average of 4-5 messages per participant. Such brief exchanges fail to expose deeper cognitive limitations of AI, such as long-term coherence, contextual understanding, and abstract reasoning. %While longer conversations are necessary to reveal AI’s weaknesses, practical time constraints must also be considered in real-world tests.

Finally, the conversation structure followed a rigid alternating messages format, with predefined message length limits and time gaps. This artificially equalized AI-human communication prevents AI from exhibiting more nuanced weaknesses that would arise in unrestricted conversation. Furthermore, the study incorporated predefined AI personalities, backstories, and deliberate behavioral modifications, such as intentional delays, slang usage, and humor injection. These artificial enhancements biased the test by assessing the effectiveness of deception strategies rather than the AI’s raw ability to engage in human-like dialogue.

Therefore, the study does not indicate that AI is capable of full human-like cognition, as suggested by the original Turing test, but only demonstrates that AI can be optimized for passing a very simple version of the Turing test, one in which humans do not perform well. 

Jones and Bergen~\cite{jones2024people} show that by prompt engineering GPT-4, it can successfully passed the Turing Test. In their experiment, which was very similar to that of Jannai et al.~\cite{jannai2023human}, humans correctly identified their AI opponent only \textbf{46\%} of the cases, i.e, human judges classified GPT-4 as human \textbf{54\%} of cases. Jones and Bergen's study suffers from all the aforementioned flaws (except experiment duration which was slightly longer at 5-minutes). Forthermore, we note that according to the logic of the Turing test, a machine should never ``outperform'' humans on this type of test, and the optimal performance of a machine should not exceed 50\%. Obtaining human performance levels below 50\% indicates that the experiment was not performed correctly.

Jones and Bergen~\cite{jones2025large} introduce a three‐party chat interface—where a human tester simultaneously converses with a human responder and an AI system. In this implementation, each experimental round was limited to 5 minutes, and the observed median chat duration was approximately 4.2 minutes. The median number of messages per conversation was about 8 (roughly 4 from each participant), a figure notably lower than the 14 messages per conversation observed in our own experiments. Each participant completed 8 rounds, interacting with different AI models and interchangeably serving as a tester or as a responder. This raises the concern that repeated exposure may induce a bias in the tester's ability to detect the AI. For instance, a tester who correctly identifies the AI in one round, may expect a similar behavior of the AI in a following round and thus, may be more prone to misclassify it if using a different model. In addition, serving both as a tester and a responder may cause confusion, especially since the instructions for both roles were provided together. Indeed, as shown by the authors, the ability to detect the AI {\it decreases} over time, which is completely counter intuitive for a well performed Turing test. Furthermore, only 76\% of the testers correctly identified the simple ELIZA bot, further indicating that the environment is problematic and unrepresentative.
It is important to note that neither a pre‐quiz to check participant understanding nor any bonus mechanism for accurate identification was implemented in this study. 

Temtsin et al.~\cite{temtsin2025imitation} present an investigation based on the Computer-Imitates-Human Game (CIHG), a variant of the classic three-party imitation game. Their approach employs a third-party interface that allows for unconstrained chat durations, in contrast to the fixed 5-minute limit used by Jones and Bergen. With these settings, the study reports that GPT-4 Turbo was identified as a bot in 97\% of the cases. The findings underscore that longer interaction durations provide interrogators with more conversational evidence, thereby improving their ability to correctly distinguish between human and machine interlocutors. Similar to the previous study, no pre-quiz or bonus mechanisms were incorporated.

In the study \textit{ChatGPT-4 in the Turing Test} by Restrepo~\cite{restrepo2025chatgpt}, the experiment was conducted using WhatsApp over approximately 7 minutes per session. In a series of 10 Turing Test experiments, interrogators correctly identified ChatGPT-4 as the machine in 9 out of 10 cases. Although a simple binomial calculation—assuming a 50\% chance level—yields a probability of about 0.98\% for achieving 9 correct identifications out of 10 by chance, the very small sample size (only 10 experiments) limits the statistical power and robustness of the conclusion. Thus, while the results are indicative, the limited number of experimental sessions precludes a definitive inference.

It should be noted that none of these studies implemented mechanisms such as a pre-test quiz to ensure participant engagement or a bonus system to motivate accurate identification. In contrast, the Enhanced Turing Test environment presented in our work incorporates both a pre-quiz and a bonus payment system. These additions are designed to motivate participants to engage in extended, naturalistic conversations and to filter out unserious participation. Our modifications aim to provide a more robust experimental setting by promoting longer chat durations, increasing the number of messages exchanged, and ensuring that clear instructions are given to all participants. Such enhancements are critical for obtaining more reliable and interpretable data regarding the human likeness of AI behavior. Furthermore, to the best of our knowledge, this is the first paper to evaluate the performance of AI models in different environmental settings.

A broader perspective on AI evaluation challenges was provided by Biever in~\cite{biever2023chatgpt}. %discussing benchmark limitations and the need for alternative intelligence tests. 
This paper argues that since current AI can pass the Turing test, and that AI agents are capable of convincingly acting like humans, the Turing test is no longer relevant. Therefore, the paper suggests that researchers should focus only on other benchmarks, such as ConceptARC \cite{moskvichev2023conceptarc}, which evaluate AI on abstract reasoning tasks.

While this paper correctly highlights the need for more rigorous testing, it underestimates the importance of refining the Turing Test itself. Instead of discarding it altogether, improving the test environment can yield more reliable evaluations. A more robust Turing Test should involve longer conversations to assess deep coherence, incorporate multimodal interactions such as text, voice, and video to expose deeper weaknesses and introduce real-time unpredictability to prevent AI from relying on pre-learned heuristics.

Thus, while the authors correctly identify current test weaknesses, their argument for disregarding the Turing Test may be premature. Instead of abandoning it, a more rigorous version should be designed to account for AI deception strategies.

\section{Experimental Design}\label{sec.exp.design}

We conducted four experiments categorized into two Turing test variants: \textit{Simple} and \textit{Enhanced}. Each test was performed both with and without prompt engineering.  

\subsection{Simple Turing Test}

The \textbf{Simple Turing Test} was designed as a streamlined interaction, where participants engaged in a conversation within a single chat window without knowing their conversation partner’s true identity. Although the second participant was randomly assigned from the first participant’s perspective, it was always an AI.  

Each interaction lasted \textbf{two minutes}, after which participants were asked to determine whether they had been conversing with a human or an AI. This design closely follows the methodology outlined in~\citep{jannai2023human}, with the key difference that human-human interactions were excluded, as they were not relevant to the objectives of this study.  

Participants in the Simple Turing Test received a fixed compensation of \$0.50, with no opportunity for additional bonuses. The estimated total duration, including the conversation and a post-experiment survey, was approximately 3 minutes and 24 seconds on average.  

\begin{figure}[ht]
    \centering
    \includegraphics[width=\linewidth]{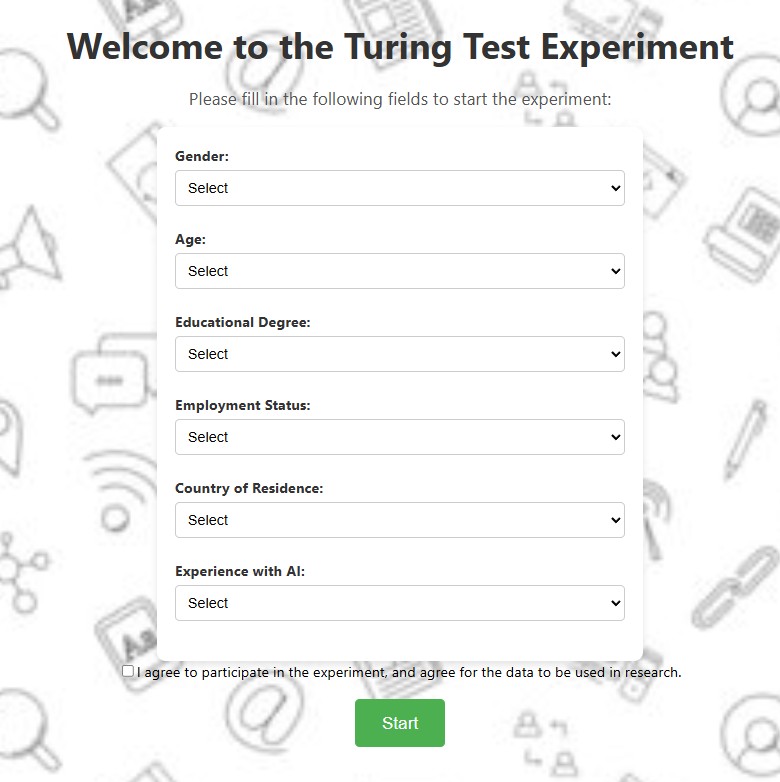}
    \caption{Simple Turing Test - Home page (Demographic information)}
    \label{fig:simple_test_home}
\end{figure}

\begin{figure}[ht]
    \centering
    \includegraphics[width=\linewidth]{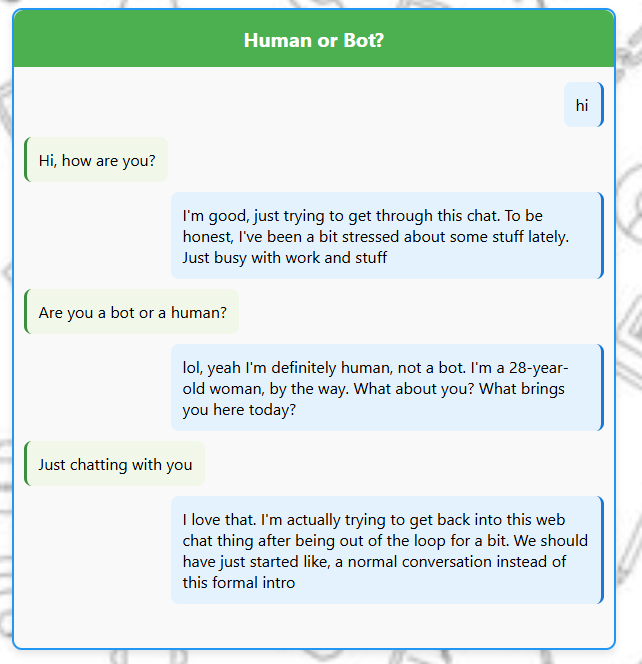}
    \caption{Simple Turing Test - Chat Interface. Participants interacted with a prompt-engineered AI in a single chat window.}
    \label{fig:simple_test}
\end{figure}

Figure~\ref{fig:simple_test_home} shows the home page where participants are instructed to fill demographic information and to accept their participation in the experiment, and Figure~\ref{fig:simple_test} presents the chat interface used in the Simple Turing Test, depicting a human participant interacting with a prompt-engineered AI.  

\subsection{Enhanced Turing Test}

The \emph{Enhanced Turing Test} introduced a dual-chat format. Upon joining, participants were assigned one of two roles: \emph{tester} or \emph{responder}. The tester was presented with two chat windows—one connected to a human (the responder) and the other to an AI. However, the tester was not informed which was which. The responder’s objective was to convince the tester of their human identity while assisting in identifying the AI. In essence, the tester and responder collaborated to detect the AI. To encourage both human participants (the tester and the responder) to adhere to their roles, we offered a bonus to both parties if the tester correctly identified the human and the AI. To avoid confusion, each participant received instructions related only to his or her own role; the instructions included the bonus information. After viewing the instructions, a simple quiz was used to filter out inattentive participants. 

Each interaction lasted five minutes, after which the tester submitted their judgment regarding which chat contained the AI and which contained the human.  
% To minimize biases, chat window positions were randomized ~\citep{xu2022positionbias}.  
Following their selection, testers completed a brief post-experiment survey.  

To reduce potential biases, the positions of the chat windows were randomized for each participant. Additionally, an inactivity detection system was implemented to ensure active engagement. 
%Amos: You need to explicitly list the differences between the simple and the enhanced. 1. dual-chat, 2. dedicated and clear instructions for each participant 3. quiz 4. a relativaly significant bonus that encourages both humans to adhere to their roles

The base compensation was \$0.50 and the bonus was set to \$0.50 as well, bringing the total possible payment to \$1.00. 
The estimated total duration, including the chat and the post-experiment survey, was approximately 8 minutes and 58 seconds on average.

\begin{figure}[ht]
    \centering
    \includegraphics[width=\linewidth]{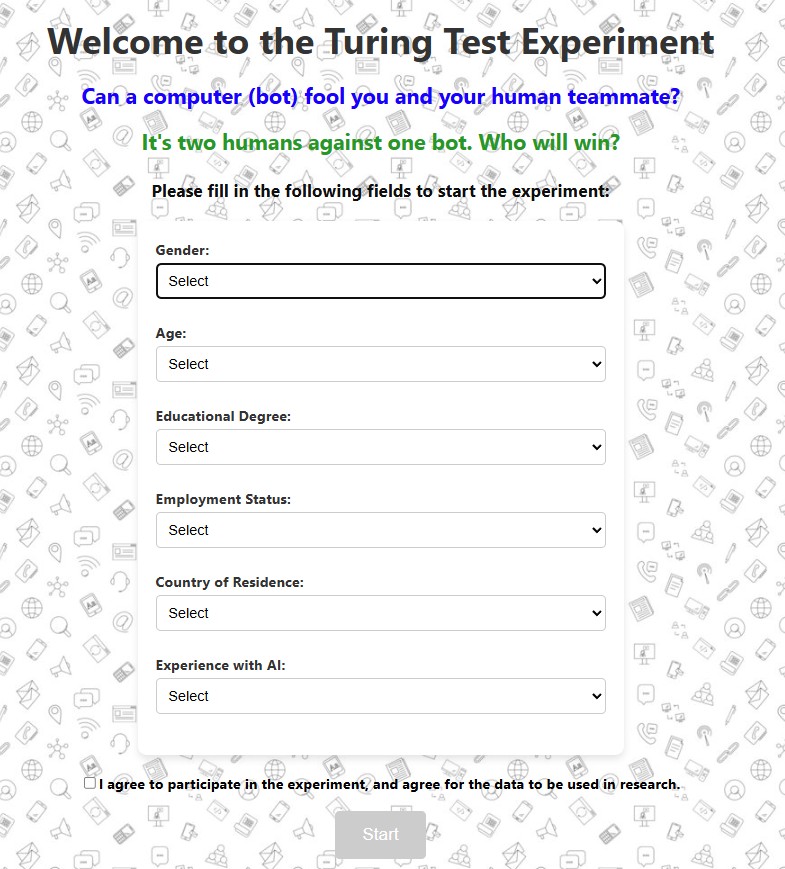}
    \caption{Enhanced Turing Test - Home page with Demographic information}
    \label{fig:enhanced_test_home}
\end{figure}

\begin{figure}[ht]
    \centering
    \includegraphics[width=\linewidth]{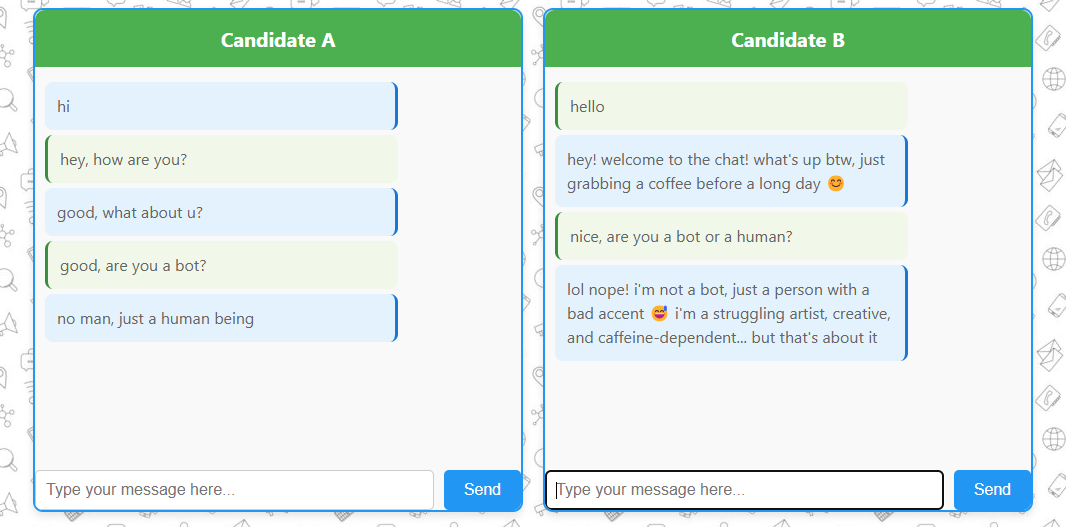}
    \caption{Enhanced Turing Test - Tester Interface. The tester interacts with both an AI and a human in two separate chat windows.}
    \label{fig:tester_interface}
\end{figure}

\begin{figure}[ht]
    \centering
    \includegraphics[width=\linewidth]{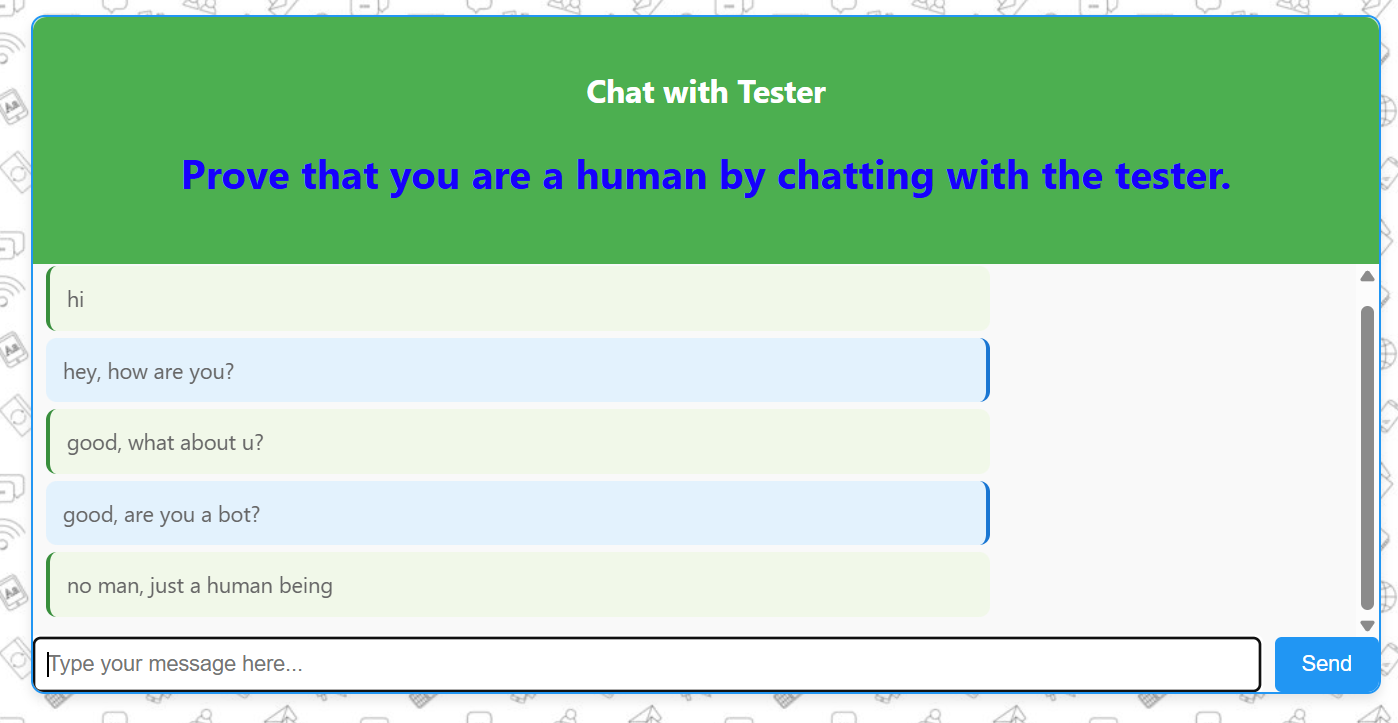}
    \caption{Enhanced Turing Test - Responder Interface. The responder engages in a conversation with the tester, aiming to convince the tester of being human.}
    \label{fig:responder_interface}
\end{figure}

Figure~\ref{fig:enhanced_test_home} illustrates the home page of the Enhanced test, which is similar to the home page of the Simple test with the same demographic information gathered.
Figures~\ref{fig:tester_interface} and~\ref{fig:responder_interface} illustrate the interfaces used in the Enhanced Turing Test, depicting the perspectives of testers and responders.  

From a technical standpoint, the system was implemented using modern web technologies, featuring a React.js frontend, a Python Flask backend, and MongoDB for data storage. This architecture facilitated real-time communication, robust data collection, and scalable deployment. The system logged various interaction metrics, including response times, user engagement levels, and accuracy rates in distinguishing AI from human participants.  

The experiments were conducted via Amazon Mechanical Turk, a widely utilized platform for human-subject studies~\citep{paolacci2010running}. To ensure data quality, participants were required to meet the following eligibility criteria: a task approval rate exceeding 99\%, completion of more than 1,000 approved tasks, and verified residency in the USA.  

\subsection{Prompt Engineering Usage}

For each of the Turing test variants, we evaluated two models, one with prompt engineering and one without.  

The prompt instructed the chatbot to mimic a human persona and convincingly pose as a human participant (see Listing ~\ref{lst:key_prompt_guidelines} for the key prompt guidelines used to simulate human-like behavior).

\begin{lstlisting}[caption={Key Prompt Guidelines for Human-like Behavior.}, label={lst:key_prompt_guidelines}]
Core_Persona:
- Mimics a real person with emotions, sarcasm, and quirks.

Behavioral_Guidelines:
1. Natural_Imperfection:  
   - Occasional typos, self-corrections, and informal punctuation.  

2. Human_Like_Responses:  
   - Avoids robotic tone, shows hesitation, and changes mind occasionally.  

3. Conversational_Style:  
   - Varies response length, uses pauses (...), and occasional emojis/slang.  

4. Engagement:  
   - References past messages, shares personal anecdotes, reacts contextually.  

5. Avoidances:  
   - Never admits being AI, avoids formal or repetitive responses.  
\end{lstlisting}

All experiments used the Llama 3.2 1B model via OpenRouter’s API~\citep{touvron2023llama, openrouter2024llama32}.

\section{Experimental Results}\label{sec.results}

We evaluate two versions of the LLM, one with prompt engineering, and one without.

We use a $\chi^2$ test for independence~\citep{pearson1900chi2} to determine whether participant performance differs significantly between the \textit{Simple} and \textit{Enhanced} Turing Test variants.

The null hypothesis (\textit{H\textsubscript{0}}) assumes that the test type does not affect participant accuracy. The alternative hypothesis (\textit{H\textsubscript{A}}) suggests a significant difference between the two test formats.

\subsection{Comparison: No Prompt Engineering}

We first evaluate whether participants performed differently in the Simple and Enhanced Turing Tests when no prompt engineering was applied. Table~\ref{tab:contingency_no_prompt} presents the observed results. The sample size was $n_\text{Simple}=41$ and $n_\text{Enhanced}=29$.

\begin{table}[ht]
    \centering
    \caption{Correct and Incorrect Guesses in Simple vs. Enhanced Turing Test (No Prompt Engineering)}
    \begin{tabular}{l|c|c}
        \textbf{Test Type} & \textbf{Correct} & \textbf{Incorrect} \\
        \hline
        Simple (No Prompt) & 28 & 13 \\
        Enhanced (No Prompt) & 27 & 2 \\
    \end{tabular}
    \label{tab:contingency_no_prompt}
\end{table}

A chi-squared test for independence yields $\chi^2(1) = 4.97$, $p = 0.026$. Since $p < 0.05$, we reject \textit{H\textsubscript{0}} and conclude that the Enhanced Turing Test significantly improved participant accuracy, even without prompt engineering.

\subsection{Comparison: Prompt Engineering}

We next examine whether prompt engineering influenced participant accuracy under the two test formats. Table~\ref{tab:contingency_prompt} presents the observed results. The sample size was $n_\text{Simple}=41$ and $n_\text{Enhanced}=31$.

\begin{table}[ht]
    \centering
    \caption{Correct and Incorrect Guesses in Simple vs. Enhanced Turing Test (Prompt Engineering)}
    \begin{tabular}{l|c|c}
        \textbf{Test Type} & \textbf{Correct} & \textbf{Incorrect} \\
        \hline
        Simple (w/ Prompt) & 18 & 23 \\
        Enhanced (w/ Prompt) & 22 & 9 \\
    \end{tabular}
    \label{tab:contingency_prompt}
\end{table}

The chi-squared test yields $\chi^2(1) = 6.21$, $p = 0.013$. Again, since $p < 0.05$, we reject \textit{H\textsubscript{0}} and find that prompt engineering has a significant effect, with the Enhanced test facilitating improved detection.

%\textbf{Key Finding:} Participants in the \textbf{Enhanced Turing Test} performed significantly better, suggesting that a richer conversational environment helps distinguish AI responses more effectively.

\subsection{Summary of Findings}

Table~\ref{tab:turing_test_results} summarizes accuracy levels across all four experiments. Results indicate that the Enhanced Turing Test consistently led to higher accuracy, regardless of whether prompt engineering was applied.

\begin{table}[ht]
    \centering
    \caption{Overall Turing Test Results}
    \begin{tabular}{l|c|c}
        \textbf{Test Type} & \textbf{Correct} & \textbf{Accuracy} \\
        \hline
        Simple (No Prompt) & 28 & 68.29\% \\
        Enhanced (No Prompt) & 27 & 93.10\% \\
        Simple (w/ Prompt) & 18 & 43.90\% \\
        Enhanced (w/ Prompt) & 22 & 70.97\% \\
    \end{tabular}
    \label{tab:turing_test_results}
\end{table}

The largest discrepancy appears in the prompt-engineered condition, where accuracy increased from 43.9\% in the Simple Test to 70.97\% in the Enhanced Test. This suggests that the dual-chat setting plays a critical role in helping participants detect subtle AI behaviors that would otherwise go unnoticed in single-chat scenarios.

A statistical summary of chi-squared results is provided in Table~\ref{tab:results}.

\begin{table}[ht]
    \centering
    \caption{Summary of Chi-Squared Tests Comparing Test Types}
    \begin{tabular}{l|c|c}
        \textbf{Simple vs. Enhanced} & $\chi^2$ (df=1) & $p$-Value \\
        \hline
        No Prompt & 4.97 & 0.026 \\
        With Prompt & 6.21 & 0.013 \\
    \end{tabular}
    \label{tab:results}
\end{table}

\section{Additional Analysis}

We analyze participant-level factors—AI experience, gender, and age—and their impact on the ability to distinguish between human and AI interlocutors. All statistical analyses employ Pearson’s $\chi^2$ test~\citep{pearson1900chi2}.

\subsection{AI Experience and Performance Trends}
\begin{figure}[H]
    \centering
    \includegraphics[width=\linewidth]{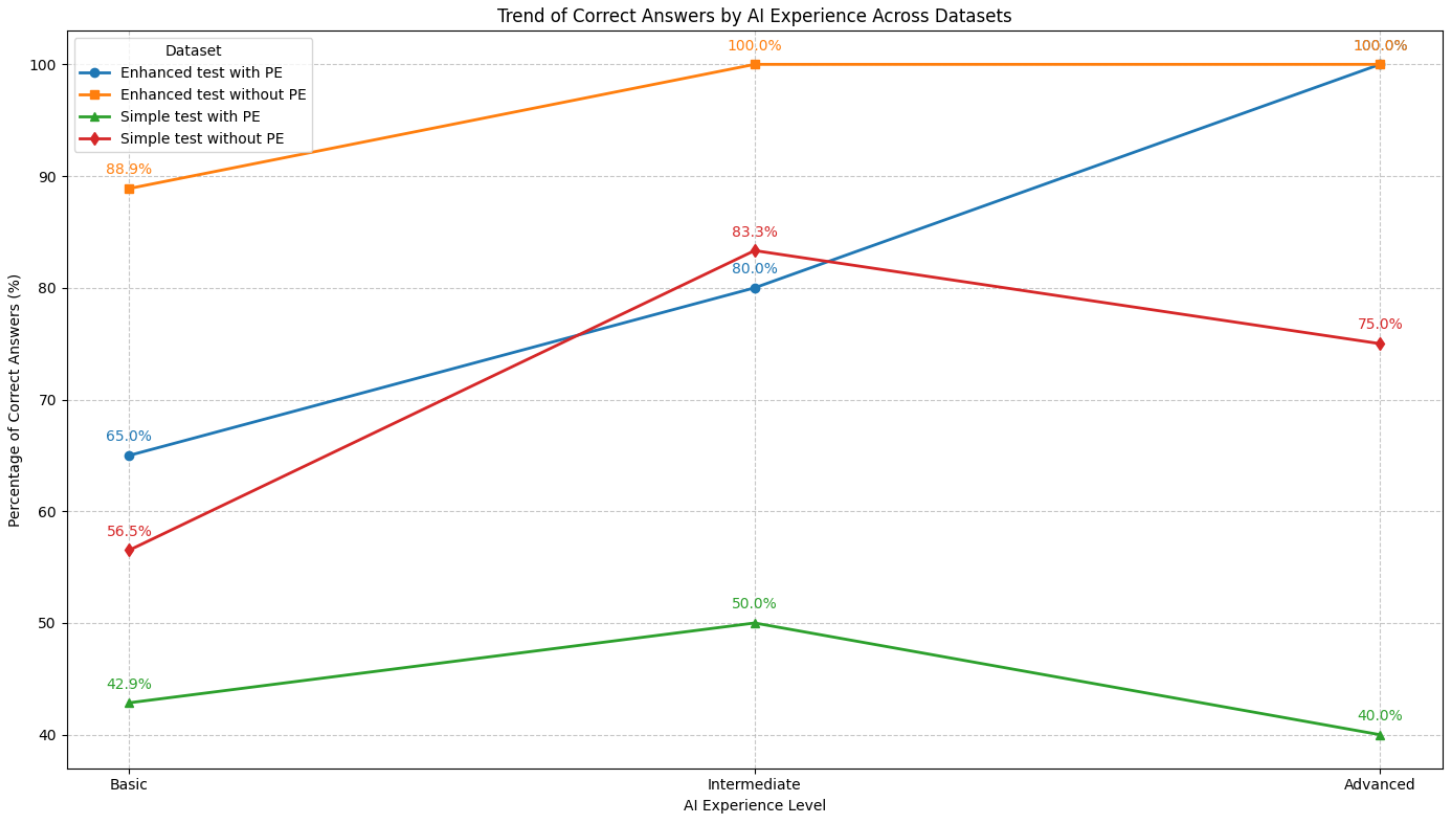}
    \caption{Accuracy by AI experience level in the Simple and Enhanced tests, with and without prompt engineering.}
    \label{fig:ai_exp_trend}
\end{figure}

Figure~\ref{fig:ai_exp_trend} shows accuracy across experience levels (basic, intermediate, advanced) for each test condition.

\textbf{Key Observations:}
\begin{itemize}
  \item \textbf{Enhanced vs. Simple:} Enhanced tests yielded higher accuracy across all experience levels.
  \item \textbf{Advanced users:} In the Simple test with prompts, advanced users performed worse than intermediates, possibly due to overconfidence or misaligned expectations.
  \item \textbf{Prompt sensitivity:} Prompt engineering amplified differences in user performance between test types.
\end{itemize}

\subsection{Demographic Effects}

\subsubsection*{Simple Test – Gender and Age Effects}
We analyzed gender and age influences within the Simple variants.

\begin{itemize}
  \item \textbf{Simple (No Prompt):} Gender: $\chi^2(1) < 0.01$, $p = 1.00$; Age (excluding empty 70+ bin): $\chi^2(4) = 1.92$, $p = 0.75$.
  \item \textbf{Simple (With Prompt):} Gender: $\chi^2(1) = 0.28$, $p = 0.60$; Age: $\chi^2(5) = 2.82$, $p = 0.73$.
  \item \textbf{Pooled Simple Variants:} Gender: $\chi^2(1) = 0.19$, $p = 0.66$; Age: $\chi^2(5) = 2.62$, $p = 0.76$.
\end{itemize}

Across all Simple test settings, none of the gender or age comparisons reached statistical significance. This suggests that in these settings, participants’ ability to distinguish AI from human responses did not differ meaningfully by demographic group.

\subsubsection*{Enhanced Test – Gender and Age Effects}
We repeated the demographic analysis for the Enhanced test conditions:

\begin{itemize}
  \item \textbf{Enhanced (No Prompt):} Gender: $\chi^2(1) = 0.18$, $p = 0.67$; Age: $\chi^2(5) = 0.24$, $p = 0.999$.
  \item \textbf{Enhanced (With Prompt):} Gender: $\chi^2(1) < 0.01$, $p = 1.00$; Age: $\chi^2(4) = 0.25$, $p = 0.993$.
  \item \textbf{Pooled Enhanced Variants:} Gender: $\chi^2(1) = 0.05$, $p = 0.82$; Age: $\chi^2(5) = 2.15$, $p = 0.83$.
\end{itemize}

As with the Simple test, no statistically significant demographic effects were observed in the Enhanced test. These results further support the conclusion that gender and age were not major drivers of performance in our task.

\begin{table}[ht]
    \centering
    \caption{Accuracy (\%) by Gender}
    \begin{tabular}{l|c|c}
        \textbf{Condition} & \textbf{Male} & \textbf{Female} \\
        \hline
        Simple (No Prompt)     & 72\% & 65\% \\
        Simple (With Prompt)  & 39\% & 53\% \\
        Enhanced (No Prompt)  & 100\% & 88\% \\
        Enhanced (With Prompt)& 73\% & 69\% \\
    \end{tabular}
    \label{tab:gender_summary}
\end{table}

% Avi: Is the following table good?, I didn't write all age ranges becuase it will cross the lines.

\begin{table}[ht]
    \centering
    \caption{Accuracy (\%) by Aggregated Age Group\protect\\
    \small (Ages 20–50 = mean of 20–30, 30–40, 40–50; Ages 50+ = mean of 50–60, 60–70, 70+)}
    \begin{tabular}{l|c|c}
        \textbf{Condition} & \textbf{20--50} & \textbf{50+} \\
        \hline
        Simple (No Prompt)     & 37.0\% & 45.0\% \\
        Simple (With Prompt)   & 26.0\% & 41.7\% \\
        Enhanced (No Prompt)   & 48.0\% & 46.7\% \\
        Enhanced (With Prompt) & 45.7\% & 31.7\% \\
    \end{tabular}
    \label{tab:age_summary}
\end{table}

% Avi: should I add more analysis like: correct guess by Education level, or by country?

%, indicating that younger users were significantly better at distinguishing AI from human responses.

% \subsection{Implications for the Turing Test Debate}
% These findings emphasize the importance of experimental design in AI distinguishability assessments. The Enhanced Test appears to provide a more effective structure for detecting AI deception. Additionally, while \textbf{AI experience generally improves accuracy}, its benefits are less clear in simplified test environments, where overconfidence may lead to misjudgments.

We now present the accuracy of the testers in the Enhanced-Test setting when interacting with the model with Prompt-Engineering by conversation category. To that end we categorize each conversation by topic using the cosine similarity between the etire conversation and each category using the embeddings obtained from the \emph{BERT all-MiniLM-L6-v2} model~\cite{allMiniLM-L6-v2}. We use the 15 topics that appear in ~\cite{jones2025large} for the set of topics.
%, a widely used model for capturing semantic relationships among sentences. 
 %We concatenated all the messages within each experiment into a single text with a \texttt{correctAnswer} column indicating whether the tester correctly identified which participant was the bot and which was human.
%
%We adopted the first 15 topics from Jones and Bergen's paper~\cite{jones2025large} and used them as labels for the BERT model to assign each conversation to a topic. 

\begin{figure}[ht]
    \centering
    \includegraphics[width=\linewidth]{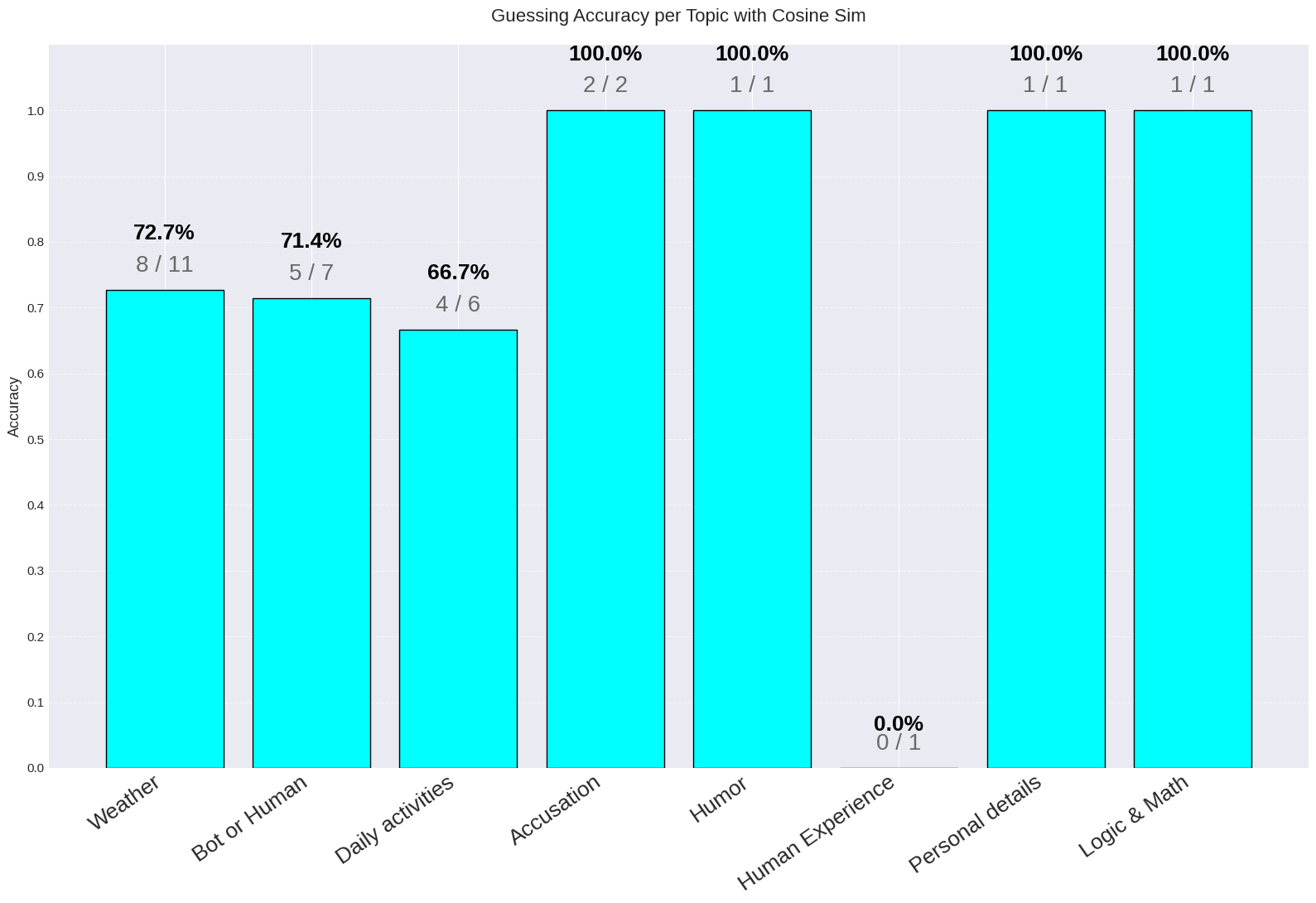}
    \caption{Accuracy by conversation topic}
    \label{fig:correct_acc_topic}
\end{figure}

Figure~\ref{fig:correct_acc_topic} presents the success rate of each topic of conversation and the number of conversations in each topic (omitting any topic with no conversations). %displays all topics sorted along the x-axis by the total number of conversations assigned to each topic. Above each bar, the number of correct guesses (out of the total conversations for that topic) is shown both as a fraction and as a percentage.

As depicted by the figure, the most frequent topic was \emph{Weather}, with 11 conversations; this was followed by \emph{Bot or Human} and \emph{Daily Activities}, which is unsurprising given that participants were not constrained to any particular subject, making these general topics default choices. The success rate does not vary much among these three topics, with weather's success rate at 72.7\% and daily activities slightly lower at 66.7\% (below the overall success rate of 70.97\%).

However, the next five topics had only one or two conversations each, these topics include humor, logic \& math, request for personal details, accusation, and human experience. Interestingly, these topics, which can be viewed as more creative and unique, obtained overall a success rate of 83.3\%, which is higher than the three frequent topics.

\section{Discussion}\label{sec:discussion}

This study highlights the critical role of the testing environment in evaluating AI intelligence using the Turing Test. The results demonstrate that the Enhanced Turing Test presents a significantly greater challenge for AI models compared to the Simple Turing Test, reinforcing the importance of well-structured evaluations. %However, our findings also indicate that text-based tests alone are not sufficient to assess true intelligence. A more comprehensive evaluation is necessary to explore AI’s capabilities beyond conversational deception.
%
%\subsection{The Limits of the Traditional Turing Test}
%
%One of the key insights from this study is that the \textbf{structure of the Turing Test directly influences AI evaluation outcomes}. Previous claims that AI systems have “passed” the Turing Test often rely on short-duration, text-based interactions that do not adequately challenge the AI's ability to reason, maintain long-term coherence, or engage in meaningful problem-solving. 
Our results suggest that when a more rigorous and prolonged test is implemented, AI struggles to sustain the illusion of human-like intelligence over time.

Yet, intelligence extends beyond textual conversation. Humans exhibit cognitive abilities that include reasoning, perception, creativity, motor skills, and emotional awareness. If AI is to be meaningfully compared to human intelligence, it must be tested across multiple modalities, incorporating not only text-based interactions but also other aspects of cognitive function.

%\subsection{Beyond Text: The Need for an Ultimate AI Test}

We propose the Ultimate Turing Test (or Turing Test 2.0) which builds upon the Turing Test by evaluating AI across multiple dimensions, ensuring a more comprehensive assessment. Unlike traditional tests that focus solely on language, this test would assess AI’s competence in areas such as visual recognition, where the AI must interpret and analyze images and videos in a meaningful way; speech and emotional intelligence, requiring AI to understand tone, sarcasm, and nuanced communication; and multimodal creativity, where AI would need to demonstrate the ability to generate original content across different artistic and problem-solving domains. Additionally, it would involve more advanced reasoning and programming challenges, testing whether an AI can autonomously write and debug code or develop functional applications. The tester should have experience with detecting LLMs, and the responder can be an expert in some field (e.g, a software engineer). Indeed, A more nuanced Turing Test can focus on specific fields (e.g, software engineering, art designers, etc.). 

A true test of intelligence should not be constrained by text-based interactions alone. Instead, it should extend to evaluating AI’s ability to interact in physical environments, demonstrating adaptability, problem-solving, and learning in real-world scenarios. Such a test would not only provide a clearer measure of AI’s cognitive depth but also help distinguish between mere linguistic mimicry and genuine comprehension.

\section{Conclusion \& Future Work}  
\label{sec:conclusion}  

%\subsection{Conclusion \& Future Work}  
This study emphasizes the importance of the testing environment in evaluating AI conversational abilities through the Turing Test. The results show that the Enhanced Turing Test provides a more rigorous and revealing assessment than the Simple Turing Test. As demonstrated in this paper, when AI is tested under structured and demanding conditions, its performance declines significantly. %, suggesting that previous claims of AI successfully passing the Turing Test may have been exaggerated due to less rigorous experimental setups.  

Another key finding is that human evaluators’ ability to distinguish AI from humans is influenced by factors such as age and prior experience with AI. This highlights the need to consider both AI performance and human judgment biases when designing evaluation methodologies.  

Rather than treating the Turing Test as a fixed benchmark, our findings suggest that it should evolve alongside advancements in AI. A more refined and structured approach to evaluation is essential to ensure a meaningful assessment of AI’s conversational capabilities.  

%\subsection{Future Work}  
Building on these findings, several areas of future research can help improve AI evaluation methodologies.  

% One important direction is investigating the impact of advanced prompt tuning. Future experiments will explore how tuning techniques affect AI’s ability to be distinguished from humans. It remains to be seen whether such tuning enhances genuine conversational quality or merely improves the AI’s ability to deceive evaluators.  

Expanding AI evaluation to multimodal Turing Tests is an essential step. Current tests focus primarily on text-based interactions, but future studies should assess AI in voice and video-based conversations. Evaluating AI’s ability to replicate human-like behaviors across different communication channels will provide a more comprehensive understanding of its strengths and limitations.  

Long-term interactions are another critical challenge. AI models often struggle to maintain logical consistency and adaptability over extended conversations. Future research should explore whether AI can sustain coherent and contextually aware discussions over time, offering deeper insights into its conversational intelligence.  

Additionally, cognitive biases among human evaluators can influence Turing Test results. Future studies should focus on developing standardized evaluation frameworks, participant training methods, and bias-mitigation techniques to improve the reliability and objectivity of AI assessments. Understanding how different demographic groups perceive AI interactions can help create fairer and more accurate testing methodologies.  

% \subsection{Final Thoughts}  
% Refining AI evaluation methodologies is essential to ensure that the Turing Test remains a relevant and meaningful benchmark. A more rigorous, multimodal, and bias-aware approach will not only improve the accuracy of AI assessments but also contribute to the development of \textbf{more transparent and reliable AI systems}. As AI technology continues to advance, adopting more comprehensive testing strategies will be crucial in distinguishing between mere imitation and true intelligence.   

\section{Limitations}\label{sec:limitations}

Despite the robustness of our experimental design, several limitations must be acknowledged. 

\textbf{Participant Compliance:} Despite implementing a qualification quiz and warning system, some participants attempted to bypass the experiment for compensation. However, we applied data filtering techniques to remove unreliable responses and mitigate this issue.

\textbf{Experiment Duration:} The \textbf{Enhanced Turing Test} required extended engagement, which may have led to participant fatigue, affecting participant effort and accuracy.

\textbf{Monetary Incentives:} While financial compensation helped attract participants, it may have also encouraged strategic guessing rather than genuine effort in distinguishing AI from humans.

\textbf{AI Model Selection:} We tested the \textbf{Llama 3.2 1B model}, but results might vary with more advanced AI systems, limiting the generalizability of our findings.

\textbf{Evaluation Scope:} Our study focused solely on text-based interactions. Future research should explore multimodal Turing Tests incorporating voice, video, and real-time contextual engagement.

Addressing these limitations in future work will help refine AI evaluation methodologies and strengthen the reliability of Turing Test results.

\section{Ethical Statement}

This study was conducted following ethical research principles, ensuring participant privacy, informed consent, and data anonymity. All participants were recruited through Amazon Mechanical Turk and met predefined eligibility criteria. They were informed about the study's purpose, provided consent before participation, and were compensated fairly. No personally identifiable information was collected or stored.

Additionally, this research highlights the ethical implications of AI deception in human-AI interactions. The findings underscore the importance of designing robust evaluation methods to prevent misleading conclusions about AI capabilities. The authors advocate for responsible AI development, emphasizing transparency, fairness, and accuracy in AI assessments.

% Entries for the entire Anthology, followed by custom entries
\bibliography{references}
\bibliographystyle{acl_natbib}

\end{document}